\documentclass[english,apjl]{emulateapj}
\usepackage{apjfonts}
\usepackage[T1]{fontenc}
\usepackage[latin1]{inputenc}
\usepackage{graphicx}
\usepackage{amssymb}

\makeatletter

\begin{document}

\shorttitle{Hot organic molecules toward IRS\,46}
\shortauthors{Fred Lahuis et al.}

\title{Hot organic molecules toward a young low-mass star:
a look at inner disk chemistry}

\author{F. Lahuis\altaffilmark{1,2},
        E.F. van Dishoeck\altaffilmark{1},
        A.C.A. Boogert\altaffilmark{3},
        K.M. Pontoppidan\altaffilmark{1,4},
        G.A. Blake\altaffilmark{4},
        C.P. Dullemond\altaffilmark{5},
        N.J. Evans II\altaffilmark{6},
        M.R. Hogerheijde\altaffilmark{1},
        J.K. J\o rgensen\altaffilmark{7},
        J.E. Kessler-Silacci\altaffilmark{6}, and
        C. Knez\altaffilmark{6}}

\altaffiltext{1}{Leiden Observatory, P.O. Box 9513, 2300 RA Leiden, The Netherlands}
\altaffiltext{2}{SRON Netherlands Institute for Space Research, P.O. Box 800, 9700 AV Groningen, The Netherlands}
\email{F.Lahuis@sron.rug.nl}
\altaffiltext{3}{Division of PMA, MS 105-24, Caltech, Pasadena, CA 91125, USA}
\altaffiltext{4}{Division of GPS, MS 150-21, Caltech, Pasadena, CA 91125, USA}
\altaffiltext{5}{Max-Planck-Institut f\"ur Astronomie, K\"onigstuhl 17, D-69117 Heidelberg, Germany}
\altaffiltext{6}{Department of Astronomy, University of Texas at Austin, 1 University Station C1400, Austin, TX 78712-0259}
\altaffiltext{7}{Harvard-Smithsonian Center for Astrophysics, 60 Garden Street, Mail Stop 42, Cambridge, MA 02138, USA}

\begin{abstract}
Spitzer Space Telescope spectra of the low mass young stellar object
(YSO) IRS\,46 ($L_{{\rm bol}}\approx0.6\ {\rm L}_\odot$) in Ophiuchus
reveal strong vibration-rotation absorption bands of gaseous
C$_2$H$_2$, HCN, and CO$_2$. This is the only source out of a sample 
of $\sim$100 YSO's that shows these features and the first
time they are seen in the spectrum of a solar-mass YSO. Analysis of
the Spitzer data combined with Keck $L-$ and $M-$band spectra gives
excitation temperatures of $\gtrsim350$\,K and abundances of
$10^{-6}-10^{-5}$ with respect to H$_{2}$, orders of magnitude higher
than those found in cold clouds. In spite of this high abundance, the
HCN $J=4-3$ line is barely detected with the James Clerk Maxwell
Telescope, indicating a source diameter less than 13\,AU. The
(sub)millimeter continuum emission and the absence of scattered light
in near-infrared images limits the mass and temperature of any remnant
collapse envelope to less than $0.01\ {\rm M}_\odot$ and 100\,K,
respectively. This excludes a hot-core type region as found in
high-mass YSO's. The most plausible origin of this hot gas rich 
in organic molecules is in the inner ($<6$\,AU radius) region of 
the disk around IRS\,46, either the disk itself or a disk wind.
A nearly edge-on 2-D disk model fits the spectral energy
distribution (SED) and gives a column of dense warm gas along the 
line of sight that is consistent with the absorption data. These 
data illustrate the unique potential of high-resolution infrared
spectroscopy to probe organic chemistry, gas temperatures and
kinematics in the planet-forming zones close to a young star.
\end{abstract}


\setcounter{footnote}{7}

\section{Introduction}
\label{sec:introduction} 
The presence of gas-rich disks around young stars is well established 
observationally and theoretically \citep*[see review by][]{greaves05}, but
comparatively little is known about their chemical structure. 
A good understanding of the chemistry is important since part of the
gases and solids in protoplanetary disks will end up in future solar
systems where they may form the basis for prebiotic species
\citep*[see reviews by][]{ehrenfreund00,markwick04}. Virtually all
observational studies of molecules other than CO have been limited to 
the outer regions, where simple organic molecules such as HCO$^+$, CN, 
HCN, and H$_2$CO have been detected at millimeter wavelengths
\citep*[e.g.][]{dutrey97,kastner97,qi03,thi04}.
Because of beam dilution, these observations cannot probe radii $<50$ AU 
from the star, which is the relevant zone for planet formation. 
High-resolution infrared (IR) spectroscopy has found CO emission 
from the warm dense gas in the inner disk region
\citep*{najita03,brittain03,blake04}, but H$_2$O is the only molecule
besides CO and H$_2$ which has been convincingly detected \citep*{carr04}.

Models of inner disk chemistry initially focused on our own primitive
solar nebula \citep*[see review by][]{prinn93} but now also consider
exosolar systems. They have grown considerably in sophistication,
including non-equilibrium chemistry, gas-solid interactions, radial
and vertical mixing, and the effects of UV radiation and X-rays from
the central star on the gas temperature and molecular abundances
\citep*[e.g.,][]{markwick02,gail02,ilgner04,glassgold04,gorti04}.
Large abundances of organic molecules like C$_2$H$_2$ and HCN are
predicted in some of these models in the inner few AU, but no
observational tests have been possible to date.

The sensitive Infrared Spectrograph (IRS) on board the Spitzer Space
Telescope opens a new window to study molecules in disks
through IR pencil-beam line-of-sight absorption spectroscopy. The
Spitzer c2d legacy program ``From Molecular Cores to Planet-Forming
Disks'' \citep*{evans03} is collecting a coherent sample of IRS
spectra of low-mass YSO's in five nearby star-forming regions. To date
more than 100 Class I and Class II sources have been observed. Of
these only one source, IRS\,46, shows strong gas-phase absorption
bands of hot molecules. 
These gas-phase bands have previously been seen only toward deeply
embedded high-mass YSO's, where they have been associated with the
inner ($\lesssim 1000$ AU) warm dense regions of the spherical
envelopes, also known as `hot cores'
\citep*{carr95,lahuis00,boonman03}. The sources studied with Spitzer
have factors of $10^4-10^5$ lower luminosity, thus limiting the
maximum temperatures and amount of warm gas in the envelope.

IRS\,46, also known as YLW16b and GY274, is part of the Ophiuchus
cloud at a distance of $\sim$125 pc \citep*{degeus89}.
It is classified as a Class I source, i.e., a protostar with a 
compact accretion disk embedded in a more extended and collapsing
envelope, based on its near- and mid-IR colors \citep*{andre94} with
$L_{{\rm bol}}\approx0.6\,{\rm L}_\odot$ \citep*{bontemps01}. 
However, the complete SED is also 
consistent with a Class II source viewed nearly edge-on, i.e., a pre-main 
sequence star with a disk but without a significant collapse envelope 
(see Sect.~\ref{sec:discussion}). It is similar to the SED of the nearly
edge-on disk CRBR 2422.8-3423 \citep*{pontoppidan05}, but IRS\,46 
has less dense foreground material. 
The favorable inclination of CRBR 2422.8-3423 allows for the
study of ices in the outer region of the circumstellar disk. IRS\,46,
also profiting from a favorable inclination, may prove to offer a
unique opportunity to directly study the gas temperature and hot
gas-phase organic chemistry in the inner disk.

\begin{figure}[t]
\includegraphics[clip,width=1.00\columnwidth]{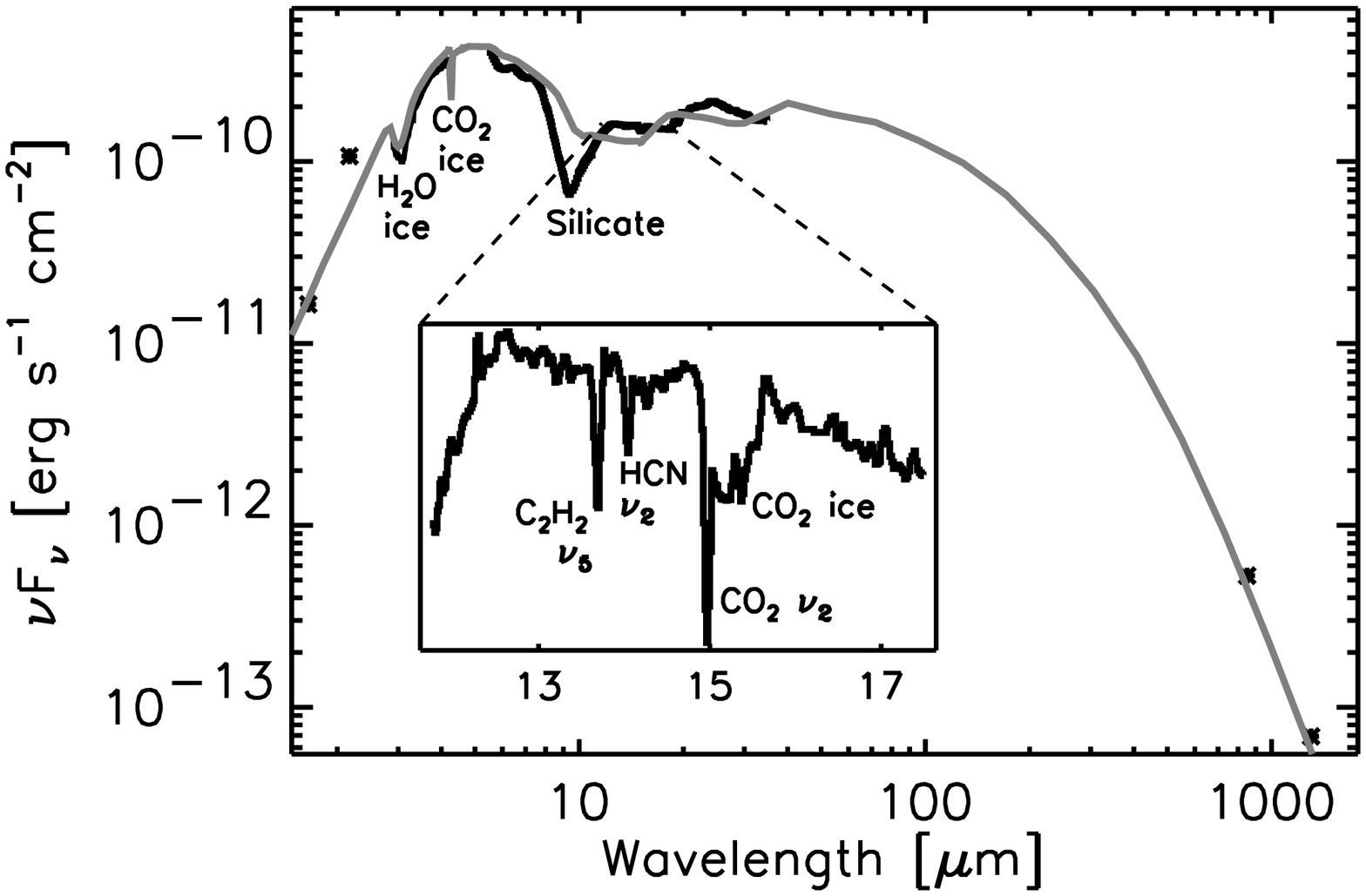}
\caption{\label{fig:sed}
Composite SED of IRS\,46 from 
1.25~$\mu$m to 1.3 mm including the full Spitzer IRS spectrum
from $10-37\,\mu$m, 2MASS $J,H,Ks$ imaging and photometry 
($F_{J}=0.33$ mJy, $F_{H}=9.03$ mJy, $F_{Ks}=77.5$ mJy), 
a VLT-ISAAC $L$-band spectrum
at $R\approx 1200$ \citep*{pontoppidan03}, an ISOCAM-CVF
$5-16\ \mu$m spectrum at $R\approx 50$,
$F_{850\ \mu{\rm m}}=150$ mJy \citep*[see][]{ridge06},
and $F_{1.3\ {\rm mm}}=28$ mJy \citep{andre94}.
Overplotted in grey is a SED disk model of a nearly edge-on self-shadowed 
flaring disk of 60 AU radius with $L_{{\rm bol}}\approx0.6\,{\rm L}_\odot$ 
and a disk mass of $\sim0.03\,{\rm M}_\odot$ seen at an
inclination of $\sim75^{\circ}$. 
The insert shows a zoom-in on the mid-IR region showing the observed 
C$_2$H$_2$, HCN, and CO$_2$ molecular absorption features and the 
CO$_2$ ice band at 15~$\mu$m.
}
\end{figure}
\begin{figure}
\includegraphics[clip,width=1.00\columnwidth]{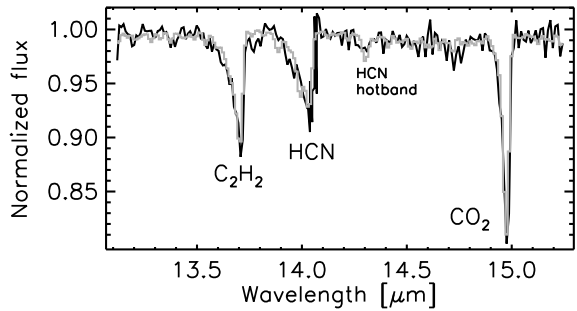}
\caption{\label{fig:c2h2_hcn_co2}
Blow-up of the IRS\,46 normalized Spitzer-IRS spectrum covering 
the ${\rm C_2H_2}$ $\nu_{5}=1-0$, ${\rm HCN}$ $\nu_2=1-0$,
and ${\rm CO_2}$ $\nu_2=1-0$ bending mode ro-vibrational absorption bands.
Included in grey is a best fit synthetic spectrum.
}
\end{figure}
\section{Observations}
IRS\,46 was observed with Spitzer-IRS in the SH ($9.9-19.6\ \mu$m) and
LH ($18.7-37.2\ \mu$m) medium-resolution modes. The observation was
positioned at ${\rm 16^h27^m29^s.4}$ $-24\degr39\arcmin16\arcsec.3$
(J2000) and acquired on 2004 August 29 as part of AOR\# 0009829888.
Low resolution SL and LL observations are scheduled as part of the
Spitzer GTO program but have not yet been taken.  Data reduction
started from the BCD images using S11.0.2 Spitzer archive data.  The
processing includes bad-pixel correction, extraction, defringing, and
order matching using the c2d analysis pipeline 
\citep[Lahuis, in preparation;][]{kessler05}.

Figure \ref{fig:sed} shows the SED composed from the full Spitzer-IRS
spectrum and complementary archival and literature data. Figure
\ref{fig:c2h2_hcn_co2} shows the part of the (normalized) IRS spectrum
revealing the ${\rm C_2H_2}$ $\nu_{5}=1-0$, ${\rm HCN}$ $\nu_2=1-0$,
and ${\rm CO_2}$ $\nu_2=1-0$ bending mode ro-vibrational absorption
bands.  Included is a best-fit synthetic spectrum (see Section
\ref{sec:results}).

To constrain the source size of the warm gas, JCMT%
\footnote{{\footnotesize The JCMT is operated by the JAC in Hilo, 
Hawaii on behalf of PPARC (UK), NRC (Canada) and NWO (Netherlands)}}
HCN $J=4-3$ and CO $J=3-2$ spectra at 354 GHz
and 345 GHz were obtained using receiver B3.
A weak HCN line with $T_{\rm MB}=0.035\,{\rm K}$ and 
${\Delta V}\approx 6\,{\rm km\,s^{-1}}$ at 
$V_{\rm LSR}=4.4 \ {\rm km\,s^{-1}}$ is detected.  
The CO $J=3-2$ line is strong, $T_{\rm MB}\approx 10$~K
at the same velocity. A small CO map around IRS\,46 reveals only a red
wing, most likely associated with the outflow from IRS~44.

\begin{figure}
\includegraphics[clip,width=1.00\columnwidth]{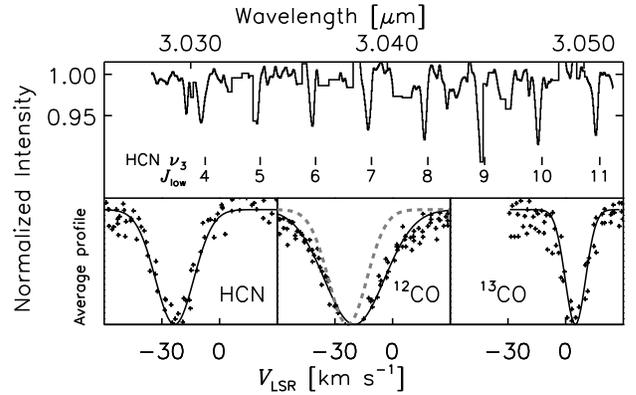}
\caption{\label{fig:hcn_co_vel} 
Top: Keck-NIRSPEC $L$-band spectrum showing HCN $\nu_3$ C-H stretching
mode absorption lines with tickmarks at the rest wavelengths. 
Bottom: HCN and CO velocity components derived from Keck-NIRSPEC 
$L$- and $M$-band spectra using absorption lines with
an atmospheric transmission $>40\,\%$.  
Plus symbols show the observed data, solid lines the fitted 
Gaussian velocity profiles.
In the $^{12}$CO profile plot the HCN profile is included 
(dashed grey line) showing that the HCN and CO lines are blue-shifted 
by a similar amount but that the CO profile is broader.}
\end{figure}
High-resolution $L$- and $M$-band spectra ($R\approx 25,000$) were
obtained with the NIRSPEC spectrometer at the Keck II telescope%
\footnote{{\footnotesize The W.M. Keck Observatory is operated as a 
scientific partnership among Caltech, the University of California 
and NASA, made possible by the W.M. Keck Foundation}}
to provide kinematic
information and confirmation of the Spitzer HCN detection through the
CO $\nu=1-0\ 4.7\ \mu$m and HCN $\nu_3=1-0\ 3.0\ \mu$m stretching mode
ro-vibrational bands (see Figure~\ref{fig:hcn_co_vel}).  The $^{13}$CO
absorption lines are unresolved at the quiescent cloud
velocity of $V_{\rm LSR}\approx 4\,{\rm km\,s^{-1}}$.  However, the
resolved $^{12}$CO ($\Delta V \approx 30\,{\rm km\,s^{-1}}$) and HCN
($\Delta V \approx 20\,{\rm km\,s^{-1}}$) absorption lines are
observed to be shifted, at $V_{\rm LSR}\approx -20\,{\rm
km\,s^{-1}}$. Of these no counterpart is observed in the HCN $J=4-3$
and CO $J=3-2$ JCMT spectra. 
We associate the HCN, C$_2$H$_2$, and CO$_2$ absorptions seen in the
Spitzer spectra with the blue-shifted components.

\section{Analysis}
\label{sec:results}
The Spitzer spectra are analyzed using a pure absorption model
assuming local thermodynamic equilibrium (LTE) excitation of the
levels at a single temperature.  The adopted method is described in
detail in \citet{lahuis00} and \citet{boonman03}, which includes
references to the molecular parameters and data used in the model.
The main fit parameters are the average temperature and integrated
column density along the line of sight for a given intrinsic line 
width, defined by the Doppler $b$-value. 
The resolved HCN $\nu_3$ lines can be the result of multiple 
unresolved components along the line of sight, so that
the intrinsic $b$ value can be smaller than the observed line width.
For small $b-$values $<2\,{\rm km\,s^{-1}}$, no good fit can be 
made to the C$_2$H$_2\,\nu_5$ and CO$_2\,\nu_2$ profiles.  
Therefore, $b$ is taken to range from 2 to 12 km s$^{-1}$.

The best fits to the ${\rm C_2H_2}\,\nu_5$, ${\rm HCN}\,\nu_2$, and
${\rm CO_2}\,\nu_2$ bands observed in the IRS spectrum give $T_{\rm
ex}$ of $\sim700$, 400, and 300 K and column densities of 3, 5, and 10
$\times 10^{16}\,{\rm cm^{-2}}$ respectively for $b\approx 5\,{\rm
km\,s^{-1}}$. The uncertainty in $b$ results in an uncertainty of
25\,\% in these values. The blue-shifted CO $\nu=1-0$ absorption band
gives $T_{\rm ex}=400\pm100$ K and $N=(2\pm 1)\times10^{18}\,{\rm
cm^{-2}}$ corresponding to a minimum H$_2$ column density of $1\times
10^{22}\,{\rm cm^{-2}}$ assuming a CO abundance of $2\times10^{-4}$
(all gas-phase carbon in CO), as appropriate for warm dense gas. The
9.7 $\mu$m silicate depth corresponds to $N_{\rm
H}=3\times10^{22}\,{\rm cm^{-2}}$ assuming a conversion factor: 
$N_{\rm H} = \tau_{9.7} \times 3.5 \times 10^{22}\,{\rm cm^{-2}}$ 
\citep*[see][]{draine03}.
X-ray observations give $N_{\rm H}=11(\pm
7)\times10^{22}\,{\rm cm^{-2}}$ \citep*{imanishi01}. 
Assuming most hydrogen is in H$_2$ and allowing for some 
foreground absorption in the latter two determinations, 
all estimates are
consistent with $N$(H$_2$)=$1\times 10^{22}$ cm$^{-2}$ within a factor
of two. The resulting abundance estimates 
are 3, 5, and 10 $\times 10^{-6}$ for ${\rm C_2H_2}$,
HCN, and CO$_2$, respectively.  The density of the gas is at least
$10^{8}\,{\rm cm^{-3}}$, required to thermalize HCN and
C$_{2}$H$_{2}$.

Additional constraints can be obtained from a combined analysis of the
HCN $\nu_2$ and $\nu_3$ bands (Figure \ref{fig:hcn_co_vel}). Assuming
the same excitation temperature for both bands, the
required column density to fit the $\nu_3$ band is higher by a factor
of 4 than that found from the $\nu_2$ band.  This suggests that
geometrical effects and emission filling in the absorption may play an
important role \citep[see][]{boonman03}. One possibility
is that the absorbing region is smaller than the
continuum emitting region, the size of which may depend on wavelength.
Similar increases may be expected for C$_2$H$_2$ and CO$_2$. 
Thus, the above cited abundances are lower
limits. The inferred HCN abundance, with respect to both H$_2$ and CO,
is up to four orders of magnitude larger than that found in cold 
interstellar clouds.

\section{Discussion}
\label{sec:discussion}
Where does this hot gas rich in organic
molecules reside? The first clue comes from the HCN submillimeter
JCMT spectrum. For a HCN column density of $> 10^{17}$ cm$^{-2}$
derived from the IR data, the $J$=4--3 pure rotational line is highly
optically thick. Thus, $T_{\rm{MB}}$ is expected to be close to the
excitation temperature of $\sim 400$ K if the emission would fill the
beam.  Although a weak emission line is observed at $V_{{\rm
LSR}}\approx4\,{\rm km\,s^{-1}}$, the broad $-20\,{\rm km\,s^{-1}}$
component is not detected with a $3\sigma$ limit of 0.02 K in a 1 km
s$^{-1}$ bin. This gives a beam dilution $\gtrsim 2\times10^{4}$
which, for the JCMT beam size of $15''$, implies a source diameter for the
hot gas of $\lesssim0.11''$ or 13\,AU diameter at the distance of
Ophiuchus. A similar limit follows from the lack of a blue wing 
on the CO $J$=3--2 line.

The second clue comes from the high temperatures and densities of the
molecular gas.  In general, temperatures of a few hundred K are found
in YSO environments only in the innermost part of envelopes or in the
inner disks.  The velocity of the hot gas provides a final clue.  The
radial velocity of IRS\,46 is unknown, and it is possible that IRS\,46
itself is at a velocity of $-20\,{\rm km\,s^{-1}}$
\citep*{doppmann05}.  More likely, however, IRS\,46 is close to the
nominal cloud velocity of $4.4\,{\rm km\,s^{-1}}$ and the hot gas is
blue-shifted by $\sim 25\,{\rm km\,s^{-1}}$.

Based on these arguments, three possibilities are considered for the
location of this hot gas: (i) the inner layer of any remnant collapse
envelope on scales $\lesssim 100$\,AU around IRS\,46; (ii) the inner
$\lesssim 10$\,AU regions of a nearly edge-on disk; and (iii) dense
hot gas at the footpoint of a wind launched from the inner disk.  The
first two options are examined through radiative transfer models to
constrain the physical parameters of the source environment.

To explore the first option, a spherically symmetric model  with
a power-law density distribution was constructed 
reproducing the SCUBA submillimeter continuum map and the
SED of IRS\,46 from mid-IR to submillimeter wavelengths
following \citet*{joergensen02}. This model has two severe problems: to
reproduce the mid-IR continuum emission, the temperature in the 
innermost envelope
cannot exceed roughly 100 K; and a $\sim0.01\,{\rm M}_\odot$
envelope (estimated from the submillimeter continuum data) produces a
significant scattering nebulosity at near IR wavelengths which is not
observed in VLT-ISAAC images \citep{pontoppidan05}.
The envelope model also cannot explain the line
widths and velocities unless IRS\,46 itself is at $-20$ km s$^{-1}$.
Thus, most of the IR and submillimeter continuum emission must 
arise from the disk around IRS\,46.

\begin{figure}[t]
\includegraphics[clip,width=1.00\columnwidth]{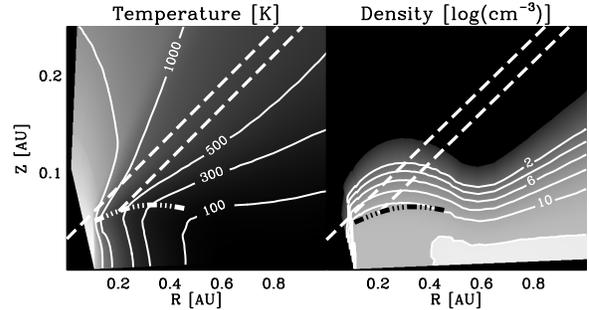}
\caption{\label{fig:disk_2d_temp_dens}
2-D distribution of the
temperature (left) and density (right) in the inner disk for the
best-fitting model to the SED. Included are the $\tau=1$ photosphere
at 14~$\mu$m (dot-dashed curves) and two lines of sight (dashed lines)
at the disk inclination of $75^{\circ}$.  Most of the molecular
absorption originates along the lines of sight toward the photosphere
in the hot ($> 300$ K) inner rim and toward the hot surface (1500 K)
of the inner rim on the opposite side of the star (see Section
\ref{sec:discussion}).}
\end{figure}
To investigate the second option, the physical structure of the
IRS\,46 disk was constrained from the observed SED in a manner similar
to CRBR 2422.8-3423 \citep{pontoppidan05} using the 2-D axisymmetric
Monte Carlo radiative transfer code of \citet*{dullemond04}. 
Figure \ref{fig:disk_2d_temp_dens} shows the temperature and density
structure of the hot inner part of the nearly edge-on self-shadowed
flaring disk and Figure \ref{fig:sed} the best fitting SED.
There is some degeneracy in the
parameters of the best-fitting model \citep[see the discussion
in][]{pontoppidan05}, but spatially resolved data are needed to
further constrain the fits. The fit to the silicate feature is
sensitive to the assumed opacities, emission from the outer disk, and
the presence of foreground absorption; these uncertainties have
little impact on the part of the disk model relevant to this work, 
i.e., the dense inner disk region.

The main continuum contribution from the disk at $3-14\,\mu$m comes from
the puffed up inner rim and inner rim wall on the far side of the star.
In the disk model, the integrated column of dense gas 
($>10^{8}\,{\rm cm^{-3}}$) toward these areas ranges from
$1-2\times10^{22}\,{\rm cm}^{-2}$ with average temperatures of 
$300-1500$ K. 
This is consistent with the H$_2$ column density and temperatures derived 
from the observations (see Section \ref{sec:results}). Indeed, more 
generally, the observed temperatures are consistent with disk models 
which explicitly calculate the gas temperature in the inner disk.
The gas temperatures may be even higher than the dust
temperatures in the upper layers \citep{glassgold04,gorti04}.
The velocity broadened CO and HCN profiles (Figure \ref{fig:hcn_co_vel})
could result from absorption in the Keplerian
inner rim at 0.1 AU ($V_{\rm kep}\simeq 70\,{\rm km\,s^{-1}}$ for a
$0.5\,{\rm M}_\odot$ star).  If IRS\,46 is at the cloud velocity of +4
km s$^{-1}$, the blue-shifted absorption implies a deviation from
Keplerian rotation in the disk plane, for example as the result of a
close binary.
A near edge-on disk explanation is consistent with a detection toward
only one in a hundred objects, since a very small fraction of sources 
should have the right orientation.

A blue-shifted velocity may also indicate that the observed absorption
features originate in a wind emanating from the inner disk.
Possibilities include a magnetocentrifugal wind (either an
X-wind launched within 0.1 AU or a disk wind launched further out), 
a photoevaporative flow or stellar wind interacting
with the upper layer and entraining molecular material 
\citep*[see e.g.][]{eisloffel00}. 
For the gas to be seen in absorption it must be comparable in size
to the inner disk region responsible for the $14\,\mu$m background,
i.e. a few AU in radius.
For a smaller region the background continuum will dominate whereas
for a much larger region line emission from the warm molecular
gas will fill in or dominate the absorption.
For a few AU region, the mass loss rate would be of order
$10^{-7}-10^{-6}$ M$_\odot$ yr$^{-1}$ assuming a density of
$\ge10^{8}$ cm$^{-3}$, a total H$_2$ column of $~10^{22}\,{\rm
cm^{-2}}$, and a flow velocity of $\sim25\,{\rm km\,s^{-1}}$. IRS\,46
does not show strong accretion signatures so a higher flow rate seems
unlikely. Quantitative predictions
in terms of velocities, column densities, densities and temperatures
of the molecular gas are needed to distinguish between the above models.

Regardless of the precise origin in the disk or disk wind, the high
inferred excitation temperatures of $400-900$ K and high abundances of
HCN and C$_2$H$_2$ of $10^{-6}-10^{-5}$ are characteristic of high
temperature chemistry.  
Hot chemistry in general is dominated by evaporation of the molecules from
the grains with subsequent gas-phase processing.  At high
temperatures, the hydrocarbon and nitrogen chemistry are enhanced
as most of the oxygen is converted into water by neutral-neutral
reactions. The abundances of molecules such as C$_{2}$H$_{2}$,
CH$_{4}$, and HCN can be increased by orders of magnitude
\citep*[e.g.][]{doty02,rodgers03} while at the same time the
formation of CO$_{2}$ is reduced as its primary formation route
through OH is blocked. Interestingly, CO$_2$ has a lower excitation
temperature in our observations.  
The most recent models of inner disk chemistry
predict enhanced abundances of HCN and C$_2$H$_2$.
In particular, \citet{markwick02} give HCN, C$_2$H$_2$, and CO$_2$ 
abundances of $10^{-6}\sim10^{-5}$ in the inner 1 AU of a protoplanetary 
disk, in good agreement with the abundances found in this work.

In summary, we present the first detection of gaseous molecular
absorption bands with Spitzer toward a solar-mass YSO, which offer
direct, unique probes of hot organic chemistry in its immediate
environment.  In addition, they provide independent constraints on the
gas temperatures and velocity patterns.  The most plausible scenario
is that the absorption originates from the inner few AU of the
circumstellar disk, perhaps at the footpoint of a disk wind.  Further
work, both observationally and theoretically, is required to prove
this, including a determination of the velocity of IRS\,46 itself and
monitoring of the infrared lines at high spectral resolution to check
for time variability of the radial velocity and/or absorbing column
along the line of sight.  This detection offers prospects for future
high spectral and spatial resolution mid-infrared and submillimeter
searches for these and other organic molecules in emission in more 
face-on disks.

\acknowledgments
We are grateful to J. Carr and J. Najita for useful discussions
and to R. Tilanus for carrying out the JCMT observations.
Astrochemistry in Leiden is supported by a Spinoza grant from NWO.
Support for this work, part of the Spitzer Legacy Science
Program, was provided by NASA through contracts 1224608, 1230779,
and 1256316 issued by the Jet Propulsion Laboratory, California
Institute of Technology, under NASA contract 1407.

\end{document}